\input{aipcheck}

\documentclass[
    ,final            
  ]
  {aipproc}

\layoutstyle{6x9}

\begin{document}

\title{The present-day galaxy population in spiral galaxies}


\author{Reynier Peletier}{
  address={Kapteyn Astronomical Institute, PO Box 800, 9700 AV Groningen}}


\begin{abstract}
Although there are many more stellar population studies of elliptical and
lenticular galaxies, studies of spiral galaxies are catching up, due to higher signal to
noise data on one hand, and better analysis methods on the other. Here I start 
by discussing
some modern methods of analyzing integrated spectra of spiral galaxies, and
comparing them with traditional methods. I then discuss some recent developments
in our understanding of the stellar content of spiral galaxies, and their
associated dust content. I discuss star formation histories, radial stellar
population gradients, and stellar populations in sigma drops. 
\end{abstract}

\maketitle


\section{Introduction}

Spiral galaxies are complicated systems, with
a number of morphological components, and several types of stellar populations. With
their, at times, violent star formation, with sometimes large amounts of
dust extinction, and with relatively low surface brightness, 
they are difficult objects to study the stellar populations
using integrated light. Studies of spiral galaxies are important, though, since
our own Milky Way belongs to this class.

Important questions to study are: When did those galaxies form? 
Did galaxies form quickly, at high $z$, as is believed for 
massive elliptical galaxies? Or did star formation take a long time.
Did bulges form first, in the same way as elliptical galaxies, before the disk?
Or did they form from disk material? For a review see Kormendy \& Kennicutt
(2004).

Most of our knowledge about the stellar populations in spiral galaxies until
recently has come
from studies of our own Galaxy, from broadband color studies (e.g.
de Jong \& van der Kruit (1994), Peletier \& Balcells 1996), from studies of ionized gas (see
Kennicutt 1998) and from a few absorption line
studies (e.g. Proctor \& Sansom 2002). In recent years, though, many more studies are
becoming available, increasing rapidly our knowledge about the stellar
populations in spiral galaxies (e.g. Ganda et al. 2007, Peletier et al. 2007,
Jablonka, Gorgas \& Goudfrooij 2007, Moorthy \& Holtzman 2006, Morelli et al.
2008, MacArthur et al. 2009). 

In this short review I will discuss a few of the issues related to stellar
populations in spiral galaxies. I will discuss their ages, their star formation
histories, and the relation between the so-called sigma-drops and their stellar
populations. I will also discuss extinction maps of a few late-type spirals. To
start, I will address methods how people can obtain 
stellar populations information from unresolved spectra in an optimal way. 
Is it better to use Lick-style indices,
or is it more efficient to directly fit the observed spectra?

\section{How to Analyze Optical Spectra?}

Stellar population synthesis of galaxies traditionally has been done using
absorption line indices of e.g. the Lick system (Worthey et al. 1994). Ages, metallicities and
abundance ratios of integrated spectra are determined by comparing indices with
stellar population models. In recent years, however, the availability of full spectral energy
distributions for the models (Vazdekis 1999, Bruzual \& Charlot 2003) has made it
possible to use the full spectral energy distribution. Some recent papers using
full spectrum fitting are Koleva et al. (2008), Koleva (2009) and MacArthur et
al. (2009). An important advantage of this method is that emission lines can be
separated very well from the absorption line spectrum,  making the assumption
that the absorption line spectrum has to resemble a linear combination of
stellar population models, convolved with an appropriate line of sight velocity
distribution (LOSVD) (see Sarzi et al. 2006 for more details). 
This means that Balmer absorption lines, which are often 
accompanied by Balmer emission lines, can be determined much more accurately
than before. But although it seems obvious
that the latter method is preferable, one can still make arguments for using
indices.

\begin{itemize} 
\item Indices are easier to publish and to compare with literature values. It is
a lot easier to compare 2 numbers than to compare 2 fits-files.
\item It is much harder to understand a fit to a continuous spectral energy
distribution than an index-index diagram. A nice example is given in Section 4
of MacArthur et al. (2009), where a misfit is seen between a galaxy spectrum and
the models in the wavelength region from 5150-5450\AA. This could be due to a flux
calibration problem, or to the fact that the galaxy has a non-solar [Mg/Fe]
abundance ratio. However, when plotting the same models in an index-index 
diagram of Mg~b vs. an Fe index, one directly sees that the misfit is caused by
a non-solar abundance ratio. 
\item The fits are ultimately limited by the quality of the models. So, if the
parameter space covered by the models does not cover the parameters of the
galaxy, unsatisfactory results are being obtained. At the moment, we don't have
a reliable set of full model spectral energy distributions with a range in age and
metallicity, but also in [Mg/Fe] abundance ratios. Models are available,
however, for the line indices of the Lick system, from Thomas et al. (2003). For
this reason some people prefer to use indices, rather than SED's. Even without
models with non-solar abundance ratios, approximations for abundance ratios can
be obtained, see e.g. Yamada et al. (2006). People,
however, are working on making full model spectral energy distributions with
variable abundance ratios, using synthetic spectral libraries, such as the one of
Coelho et al. (2005). 
\end{itemize}

To me the most important achievement of the full SED fitting method is that it
is possible to obtain reliable star formation histories with multiple bursts.
This requires a large wavelength range, an accurate treatment of the effects of
velocity broadening, and excellent single age/metallicity (SSP) 
stellar population 
models. Koleva (2009) showed that she needs a combination of up to 4 SSP's per
galaxy to fit the spectra of a sample of dwarf elliptical galaxies. Similar star
formation histories are obtained if she fits these spectra using a completely
different and independent way, using STECKMAP (Ocvirk et al. 2006). Similar
quality results are obtained by MacArthur et al. (2009) for a sample of spiral
galaxies. Using indices, on the other hand, it is already difficult to determine whether the galaxy
contains one or two SSP's (e.g. Serra \& Trager 2007).


\section{Star Formation Histories and Radial Stellar Population Distributions}

Thomas \& Davies (2006) claim that small bulges of galaxies, with central
velocity dispersions smaller than 100 km/s, have extremely low ages, sometimes
smaller than 2 Gyr. Here it should be understood that they mean
luminosity-weighted ages. On the basis of this they conclude that these bulges
have been rejuvenated with at least 10-30\% of the mass having recently formed.
This result is questioned by MacArthur et al. (2009), who, with their more
accurate stellar population modeling from full SED fitting, find a considerable
old stellar population in all of their 8 galaxies ($\ge$ 70\% in mass, bust
mostly more than 80 or 90\%. For galaxies with $\sigma_c$ $\sim$ 60 km/s, the latter
find an average luminosity-weighted age of 6 Gyr, much higher than Thomas et al.
(2005). Ganda et al. (2007), for a sample of Sb-Sd galaxies, 
modeled their Lick-indices in the context of an
exponential star formation history. Converting their best-fit e-folding time
scales into average ages, MacArthur et al. (2009) showed that they correspond also to
luminosity-weighted ages of $\sim$ 6 Gyr at $\sigma_c$ $\sim$ 60 km/s.

Independent evidence of a large range in ages in bulges of spiral galaxies is
given by the SAURON results of Peletier et al. (2007). In their sample of Sa and
Sab galaxies they find objects for which the bulges are uniformly old, objects with young
stellar populations in a small central region, others with young stars across
the whole bulge, and bulges that contains rings forming large amounts of stars.
Current wisdom in the literature is that there are two modes of star formation 
(Kennicutt 1998). The first manifests itself by a strong relation between the 
morphological type and the amount of star formation, as measured from H$\alpha$.
This causes the average star formation rate to increase monotonically with
morphological T-type. The second mode can be found in the circumnuclear regions of many spiral galaxies, which harbor
luminous star-forming regions. The physical conditions in
the circumnuclear star-forming discs are distinct in many respects from the more extended
star-forming discs of spiral galaxies. The circumnuclear star formation is especially distinctive
in terms of the absolute range in SFRs, the much higher spatial concentrations of gas and stars,
and its burst-like nature (in luminous systems) (Kennicutt 1998).

Although the picture described above was derived primarily from H$\alpha$
images,  Peletier et al. (2007) see exactly the behavior described  above.  The
galaxies, all early-type spirals, and most of them of type Sa, show a much
larger range in age than ellipticals and S0s, derived both from the index-index
diagrams and from the Mg (or H$\beta$ )- $\sigma$ diagram. Also spatially, one sees
that the younger stellar populations are concentrated near the center or in
annuli suggestive of resonance rings (e.g. Byrd et al. 1994). The picture in
these early-type spirals is
consistent with all star formation taking place in the disc, and close to the
center. In a different study, of a sample of edge-on S0-Sc galaxies, Jablonka et
al. (2007), from minor axis spectra, find that outside the plane of the galaxy, 
which they cannot
investigate because of dust extinction, the stellar populations are very similar
to those in ellipticals and S0's, with metallicity decreasing outward, and ages
somewhat younger near the center of the galaxy. Putting together all the
evidence from the different samples, it seems that the central regions of spiral
galaxies consist of a spheroidal component (a classical bulge), 
with elliptical-like stellar populations, and a disk-like component (a
pseudo-bulge), with a lower scale height, and generally younger stellar 
populations. 

A galaxy population that has not been studied often has been the stellar
populations in galactic bars. Martin \& Roy (1994) concluded from ionized gas 
studies that   global abundance gradients of spiral galaxies with bars are in
general shallower than gradients of normal galaxies. No studies of gradients of
absorption lines in reasonably-sized samples have been published until the
recent work of P\'erez et al. (2008), who show some interesting results:
They find three different types of bars according to their metallicity and
age distribution as a function of radius:  
1) Bars with negative metallicity gradients.
These show young/intermediate luminosity-weighted average ages ($<$~2~Gyr),  
and have amongst the
lowest central stellar velocity dispersions. 2) Bars with zero
metallicity gradients. These galaxies, without any gradient in
their metallicity distribution along the bar, have negative age gradients
(i.e younger populations at the end of the bar).  
3)  Bars with positive metallicity
gradients, i.e. that are more metal rich at the ends of the bar. 
These galaxies are
predominantly those with higher velocity dispersions and highest average ages.  

\section{Dust Extinction in Late-Type Spirals}

When determining stellar populations, it is often also important to determine
the amount of extinction in spiral galaxies. Given the recent renewed interest 
in this subject (Driver et al. 2007, Graham \& Worley 2008) Ganda et al. (2009)
decided to calculate central extinction maps of a number of Sb-Sd galaxies by
combining SAURON Mg~b absorption line maps, which are basically independent of
extinction, with broad band HST $V-H$ color maps. The combination of color and
line strength gives the color excess E$_{V-H}$ in a model-independent way,
which can easily be converted to A$_V$. This method is in principle very
powerful, but has not been applied much in the literature, because of the lack
of well-calibrated line strength data.
They find that the central A$_V$ ranges from 0 to 2 mag, with many galaxies
being optically thick in the optical. The  $V-H$ profiles show a lot of
structure, mostly due to extinction, with much larger  gradients than are
generally displayed by early-type galaxies. One can conclude from this that in
general color gradients cannot be used to determine stellar population
gradients. I show the extinction maps of 5 of
their galaxies in Figure 1.

\begin{figure}\centering
{\includegraphics[height=20truecm]{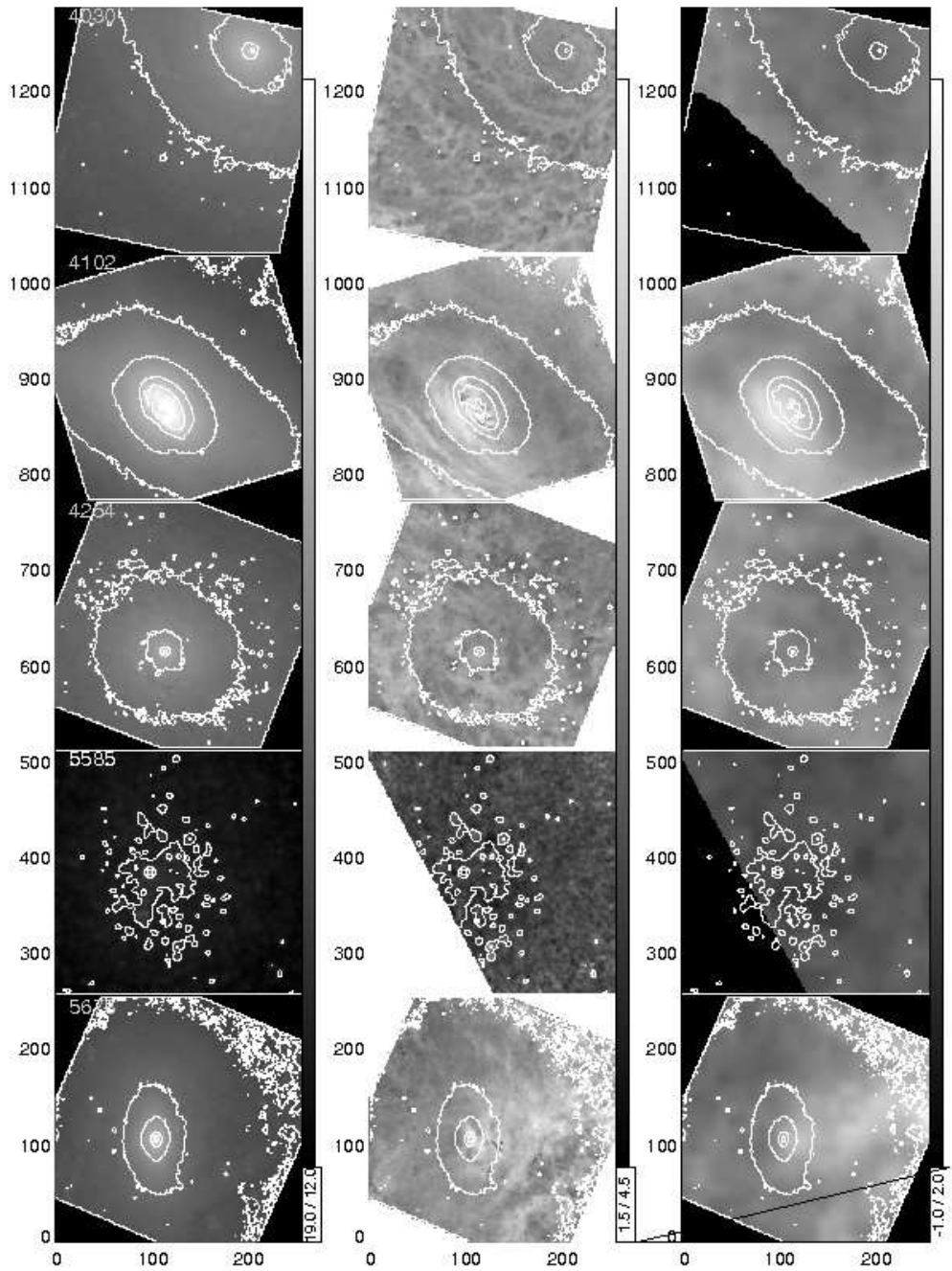}}
\caption{Extinction distribution for 5 of the 10 galaxies from Ganda et al.
(2009) with HST in
F606W and F160W. Left: surface brightness distribution in H. Middle: calibrated
V-H color maps. Right: extinction maps from $V-H$ and Mg~b. The ranges of the
plots are shown in the lower right of each column of plots.}
\label{fig22ch4}
\end{figure}

\section{Stellar Populations and Sigma Drops}

Galaxies with local velocity dispersion minima in their center (so-called
sigma-drops) are found to be increasingly common. Current wisdom is, that they
are relatively cold stellar disks that formed through secular processes in the
disk. In a review in 2006, Emsellem (2006) gives a list of 25 spiral galaxies
with sigma-drops. In the SAURON-papers of Ganda et al. (2006) and Peletier et
al. (2007) another 21 are given out of a total combined sample of 42, indicating
that about 50\% of all spirals contains sigma-drops, first detected by Bottema
(1989). As a comparison, Chung \& Bureau (2004)
revealed a sigma drop frequency of about 40\% in a sample of 30 nearly edge-on
S0-Sbc galaxies, which is consistent with the previous number of 50\%. The 
2-dimensional data of SAURON
clearly confirm that the  $\sigma$ drops are caused by central discs. In the
2-dimensional area where $\sigma$ is lower, the stars are always rotating faster than
at larger radii. 

The fact that we observe so many sigma-drop galaxies must imply that these disks
are long-lived. These discs are formed as dynamically cold systems, and slowly
heat up.  As long as they are cold,  and are responsible for a significant
fraction of the light, they will produce $\sigma$ drops at any inclination, if
they are dominating the light.  This interpretation is consistent with  N-body
and SPH simulations of, e.g.,  Wozniak \& Champavert (2006), who form discs
from  gas inflow towards the central regions of the galaxy and subsequent star
formation. 

The SAURON data offer a nice possibility to check whether these central disks
are indeed long-lived, by measuring their stellar population ages. 
In Figure 10 of Peletier et al. (2007) the square root of the quadratic 
difference between the maximum and 
central velocity dispersion is plotted. This is a measure of the size of the 
central dip. If no dip is present this parameter is equal to zero.
Here we see that, although the
fraction of $\sigma$-drops for young galaxies is larger, very old central discs
exist as well. They conclude that indeed central discs can survive long. 
McDermid et al. (2006) shows the age distribution of kinematically-decoupled
cores (KDCs) in the SAURON/OASIS sample of early-type galaxies. 
All large-scale KDCs ($\sim$1 kpc), present in slow-rotating galaxies, appear
old, while most small-scale KDCs ($\le$ 300 pc), present in fast-rotating
galaxies, appear young. Their interpretation is that they are also discs, which
like any stellar population will slowly fade, until the surface brightness is
low enough to be totally overcome by the main galaxy body. The sigma drops of
spiral galaxies discussed here are extended (size $>$ 5$''$), and 
mostly older than 1 Gyr. They
might be similar to objects such as the central discs in giant ellipticals (for
more discussion see McDermid et al. 2006).

\section{Summary}

In this short review I have discussed a number of recent developments
in our understanding of the stellar content of spiral galaxies, and their
associated dust content. Here I briefly state the conclusions.

\begin{itemize}
\item Instead of using Lick-style indices, more and more people using full SED
fitting methods to analyze galaxy spectra. These methods are becoming increasingly 
mature, and are clearly superior to the old methods for most applications.
Using full SED fitting accurate detailed star formation histories can be
determined. One should continue to realize that problems with the current
stellar population models, such as the non-availability of SSP-models with
non-solar abundance ratios, also apply to the SED fitting method.
\item The central regions of spiral galaxies show several components: a
spheroidal component containing similar stellar populations as in elliptical
galaxies and S0's, and a disk-like component, with complex star formation
histories. About 50\% of the galaxies contain central inner, fast rotating disks, 
that cause sigma-drops. The stellar populations in these inner disks can be both
old or young.
\item The comprehensive study by P\'erez et al. of stellar populations along
bars shows that there are 3 kinds of bars: those with negative metallicity
gradients, those with positive metallicity gradients, and those with zero
metallicity gradients, along which the age becomes younger towards the end of
the bar.
\item Using calibrated HST color maps and SAURON Mg~b absorption line maps,
Ganda et al. (2009) have made extinction maps that are almost model-independent.
They show that Sb-d galaxies have central A$_V$ that ranges from 0 to 2 mag, 
with many galaxies being optically thick in the optical. The  $V-H$ profiles 
show a lot of
structure, mostly due to extinction, with much larger gradients than can be
attributed to stellar population gradients.
\end{itemize}

\section*{Acknowledgements}

I thank by collaborators for all the hard work that they have put in, to produce
interesting science. I thank the conference organizers for providing a meeting
in a nice atmosphere.






\end{document}